\documentclass{iaus}
\usepackage{graphicx}

\def \kms {${\rm{km}\,\rm{s}^{-1}}$}

\def \ha   {H$\alpha$}

\def \hg   {H$\gamma$}

\def \afe  {[$\alpha$/Fe]}
\def \zh  {[Z/H]}

\title[Age and metallicities of faint red galaxies] 
{Ages and metallicities of faint red galaxies in the Shapley Supercluster}

\author[Smith, Lucey \& Hudson]   
{Russell J. Smith$^1$%
,
John R. Lucey$^1$ \and Michael J. Hudson$^2$}
\affiliation{$^1$Department of Physics, University of Durham, United Kingdom\\[\affilskip]
$^2$Department of Physics and Asronomy, University of Waterloo, Canada}

\pubyear{2007}
\volume{245}  
\pagerange{???--???}
\date{?? and in revised form ??}
\jname{Proceedings Title IAU Symposium}

\editors{M. Bureau, E. Athanassoula, and B. Barbuy, eds.}
\begin{document}

\maketitle

\begin{abstract}
We present results on the stellar populations of 232 quiescent galaxies in the 
Shapley Supercluster, based on spectroscopy from the AAOmega spectrograph at the 
AAT. The key characteristic of this survey is its 
coverage of many low-luminosity objects ($\sigma\sim{}50$\,\kms), 
with high signal-to-noise ($\sim$45\,\AA$^{-1}$). Balmer-line age estimates 
are recovered with $\sim$25\% precision even for the faintest sample members. 
We summarize the observations and absorption line data, and present correlations 
of derived ages and metallicities with mass and luminosity. We highlight the strong 
correlation between age and $\alpha$-element abundance ratio, and the anti-correlation of 
age and metallicity at fixed mass, which is shown to extend into the low-luminosity regime. 
\end{abstract}

\firstsection 
\section{Introduction}

The evolution of galaxy populations, subsequent to the peak of cosmic star formation at $z\sim2$, has been 
driven by a progressive extinguishing of activity as galaxies lose or exhaust their gas. Recent surveys of 
galaxies at redshifts up to $z\sim1$ (e.g. Bell et al. 2004; Faber et al. 2007)  reveal the corresponding 
growth in stellar mass on the ``Red Sequence'' of quiescent galaxies. A theme which
emerges from such work is the ```downsizing'' trend, whereby
star-formation appears to be shut off at high redshifts in the most massive galaxies, while 
lower-mass galaxies are ``quenched'' only at more recent epochs.

Evidence for downsizing is also found in cluster galaxy populations. 
Distant clusters are deficient in faint Red-Sequence members, relative to 
local clusters, suggesting that low-mass galaxies continued to form stars well after the giant ellipticals 
were quenched (e.g.  De Lucia et al. 2007; Stott et al. 2007). 
There is also evolution in the typical mass of ``post-starburst'' (post-quenching?) galaxies, with more
massive galaxies passing through this short-lived phase at higher redshifts than the low-mass galaxies. 
(Tran et al. 2003). 

An alternative to studying galaxy populations at higher redshifts is to infer the quenching history
from detailed spectroscopic analyses of {\it today's} Red Sequence galaxies. In particular, comparison 
of observed metal and Balmer line strengths against model predictions can yield estimates of stellar population
``age'', weighted towards the most recent star-formation episode. Until recently, high signal-to-noise (S/N) absorption-line 
data were limited to fairly small samples of galaxies, and generally to giant elliptical galaxies (e.g. 
Kuntschner et al. 2001). Samples based on much larger spectroscopic surveys (e.g. Nelan et al. 2005; 
Bernardi et al. 2006; Graves et al. 2007) generally do not have sufficient S/N to measure ages for individual 
galaxies, except at the highest luminosities. Nonetheless, these surveys do recover an {\it average} trend of younger ages for galaxies with 
lower $\sigma$, which is the expected signature of downsizing in the low-redshift fossil record. 

Downsizing implies that today's {\it faint} Red Sequence galaxies are of crucial importance to 
understanding the evolution of galaxy populations at $z<1$. 
Some early progress in measuring galaxy ages in the low-luminosity regime was made by Mobasher et al. (2001) and 
Caldwell, Rose \& Concannon (2003). Here, we present new results from a comprehensive spectroscopic survey of 
galaxies in the Shapley Supercluster ($z=0.045$) using the AAOmega spectrograph at the Anglo-Australian Telescope. 
Our study is unique in its combination 
of wide luminosity baseline coverage, large sample size, and high S/N maintained even at low luminosity. 

Section~\ref{sec:obs} of this paper summarizes the material published as Smith, Lucey \& Hudson (2007a). Section~\ref{sec:ages} 
presents further analysis, to appear in subsequent papers (Smith, Lucey \& Hudson, 2007b,c, in preparation).

\section{AAOmega observations and the index-$\sigma$ relations}\label{sec:obs}

In April 2006, we obtained spectra for a total of 416 galaxies in the Shapley Supercluster, forming a magnitude-limited ($R<18$) 
sample from the photometric catalogue of the NOAO Fundamental Plane Survey (NFPS, Smith et al 2004). 
The sample covers the central 40$\times$40 arcmin$^2$ region in each
of the three clusters Abell 3556, Abell 3558 and Abell 3562.
Two AAOmega fibre configurations were employed, with  $\sim$8 hours total integration for each configuration. 
The spectra cover the range 3700--5800\,\AA\ 
at a resolution of 3.5\,\AA\ FWHM in the blue spectrograph, and 5800--7300\,\AA\ at 1.9\,\AA\ FWHM in the red arm. 
The median S/N for faint galaxies ($\sigma<100$\,\kms) is $\sim$45 per \AA, 
(at rest-frame 4400--5400\,\AA). For galaxies with $\sigma>100$\,\kms, the median S/N is $\sim$90 per \AA. 

Redshifts and velocity dispersions were measured using standard cross-correlation methods. Emission line
equivalent widths were measured after removing a best-fiting stellar continuum model. The Lick absorption line indices
(Trager et al. 1998), were measured on the blue-arm spectra, and converted to the Lick system using
resolution transformations based on model SSP spectra, to avoid smoothing the observed spectra. 

\begin{figure*}
\includegraphics[angle=0,width=135mm]{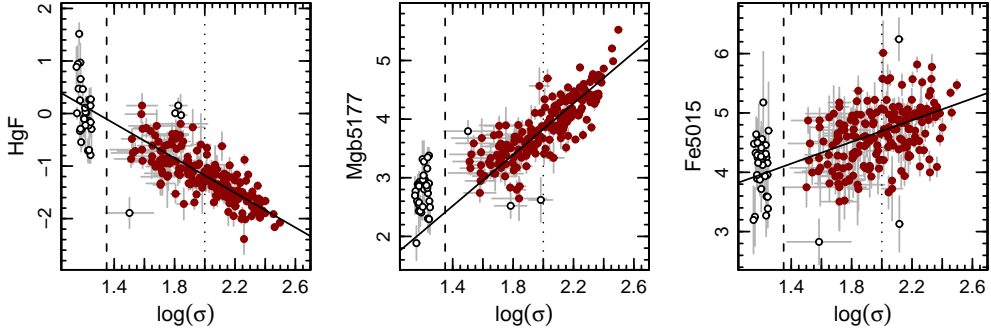}
\vskip -2mm
\caption{
The index-$\sigma$ relations for selected indices.
Objects to the left of the dashed line have velocity dispersions consistent with zero, and are not used in fitting the relations.
Open symbols represent these and also other galaxies excluded from the fit by an iterative outlier rejection.
The dotted line at $\sigma=100$\,\kms\ shows the typical low-mass limit of many other studies.
}\label{fig:ixsigs}
\end{figure*}

Our stellar population analysis is restricted to supercluster members using the measured redshift, 
and to quiescent galaxies defined by their \ha\ emission,  EW(\ha) $<0.5$\,\AA, yielding a sample of 232 galaxies. 
Figure~\ref{fig:ixsigs} shows the correlations of selected absorption line indices with velocity dispersion for this 
sample. Note that $\sim$30 galaxies have measured velocity dispersions consistent with zero and are shown at an arbitrary $\sigma$ at the left
of the figure. Index$-\sigma$ slopes are fitted to the $\sim$200 galaxies with measured $\sigma$. 

The slopes of the index$-\sigma$ relations can be used to infer the scaling relations followed by the stellar
population parameters (e.g. Nelan et al. 2005; Smith et al. 2006). For the Thomas et al. (2003) models, the
derived parameters are age, total metallicity, \zh\ and $\alpha$-element abundance ratio, \afe, estimated by 
comparison to predictions for simple stellar populations (SSPs). 
Fitting for the slopes of nine index$-\sigma$ relations, 
in comparison to the these models, we recover:
$
{\rm Age}\,\propto\sigma^{0.52\pm0.10},\ 
{\rm Z/H}\,\propto\sigma^{0.34\pm0.07},\ 
{\rm and\ }
\alpha/{\rm Fe}\,\propto\sigma^{0.23\pm0.06}.
$
The quoted errors include a (dominant) contribution from sytematic errors, estimated using a large set of different
indices in the fitting process. 

The age--$\sigma$ scaling relation is consistent with that obtained by Nelan et al. (2005) 
from similar analysis of the NFPS, and implies that even if the most massive red galaxies 
formed at very high redshift, much of the faint red population became quiescent only at recent epochs ($z<0.5$). 
As shown in Smith et al. (2007a), the implied deficit
of faint red galaxies as a function of redshift agrees {\it quantitatively} with the observed evolution in the
Red Sequence luminosity function (Stott et al. 2007; De Lucia et al. 2007). 

\section{The Age--Metallicity--Mass relations}\label{sec:ages}

\begin{figure*}
\includegraphics[angle=0,width=135mm]{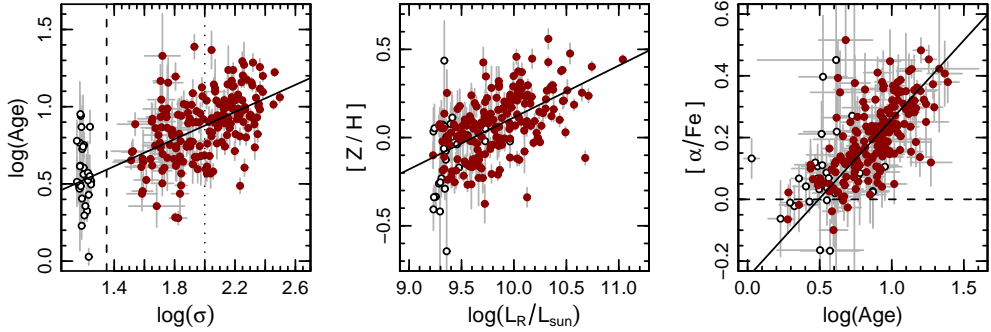}
\vskip -2mm
\caption{Left: Ages versus velocity dispersion. Centre: total metallicity versus luminosity. Right: \afe\ versus age. The
SSP-equivalent parameters were derived by comparison to models of Thomas et al. (2003). Galaxies with unresolved velocity dispersions are shown 
by open symbols and placed at arbitrary position in the left panel. 
}\label{fig:agesig}
\end{figure*}

To move beyond the average scaling relations, we use stellar population models to determine the
values of age, \zh\ and \afe\ which reproduce a non-degenerate set of observed indices, 
for each galaxy individually.
The results shown here are derived from fitting to the models of Thomas et al. (2003), using indices 
\hg$_{\rm F}$, Fe5015 and Mgb5177. For the {\it faintest} galaxies in the sample ($L_R \sim 10^{9.5} L_\odot$), the median 
formal errors are 25\% in age, 0.08 dex in Z/H and $\alpha$/Fe.
Some key correlations of the derived parameters are shown in Figure~\ref{fig:agesig}.
Intriguingly, we find that although age is correlated more tightly with $\sigma$ than 
with luminosity, metallicity is better correlated with luminosity. 
The abundance ratio \afe\ is well-correlated with $\sigma$ and also follows a tight age--\afe\ relation.
The latter suggests that ``younger'' galaxies formed stars over more extended periods, 
broadly consistent with a quenching scenario rather than formation in a single rapid burst.

\begin{figure}
\includegraphics[angle=0,width=135mm]{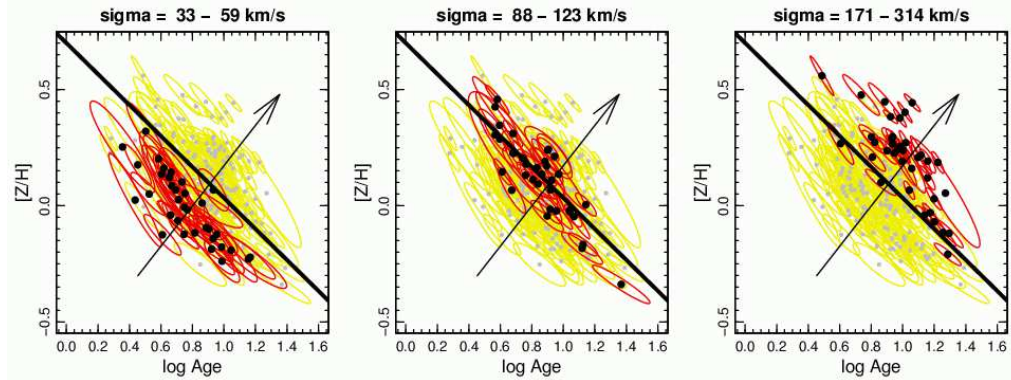}
\vskip -2mm
\caption{The Age--Metallicity--Mass relations. The black points and red ellipses show the estimated and 1$\sigma$ confidence
intervals for galaxies in each of three intervals in velocity dispersion. The rest of the sample is shown in grey/yellow for reference.
The arrow indicates the direction in which the median age and metallicity move with increasing mass.
The heavy black line (same in all panels) indicates the direction of the age--metallicity degeneracy:
movement parallel to this line generates no change in galaxy broadband colours.
}\label{fig:ammr}
\end{figure}

Figure~\ref{fig:ammr} shows the relationship between age and metallicity for three intervals in $\sigma$.
As expected from the previous figure, both age and \zh\ increase with increasing $\sigma$, on average. 
However, at {\it fixed} $\sigma$, they are anti-correlated: galaxies
which are younger than average for their mass are also more metal-rich than average. The sample populates
a plane (the ``Z-plane'') described by
$
[{\rm Z}/{\rm H}] = 0.66\pm0.04 \log\sigma - 0.68\pm0.04 \log {\rm Age}  - 0.64\pm0.08 ,
$
with scatter $\sim$0.1\,dex in \zh. The coefficients are consistent with those obtained by 
Trager et al. (2000) for a smaller sample, mostly comprising giant elliptical galaxies. Our
results demonstrate the continuity of this plane into the low-luminosity regime.

The age coefficient in the Z-plane is such that the distribution of galaxies is 
approximately aligned with the age--metallicity degeneracy, so that the reddening effect 
of higher metallicity compensates for the bluer colours expected of younger populations.
The tightness of the tight colour--magnitude sequence is partly a consequence of this ``conspiracy''.
Note that although the error ellipses are oriented parallel to the intrinsic dispersion, 
the measurement errors can account for only $\sim$20\% of the observed variance along the plane.

The existence and slope of the age--metallicity anti-correlation at fixed mass is 
crucial to understanding the star-formation history of the galaxies prior to 
quenching. In Smith, Lucey \& Hudson (2007c, in preparation), we describe a simple 
self-enrichment model for the pre-cursors of today's Red Sequence galaxies, and show that the
rate of metallicity growth required to fit the Z-plane agrees {\it quantitatively} with the 
evolution observed in the metallicity--mass relation for star-forming field galaxies 
(e.g. from Erb et al. 2006).

\begin{acknowledgments}

RJS is supported under the STFC rolling grant PP/C501568/1 
``Extragalactic Astronomy and Cosmology at Durham 2005--2010''.

\end{acknowledgments}

\end{document}